\documentstyle[aps,prc,floats,epsfig]{revtex}

\def\dag{$d(\alpha,\gamma)^6{\rm Li}$ }
\def\h2{$^2{\rm H} $}
\def\he4{$^4{\rm He} $}
\def\li6{$^6{\rm Li} $}

\begin{document} 
%@twocolumnfalse\endcsname
\draft
\title{A six-body calculation of the alpha-deuteron radiative capture cross
section}
\author{K. M. Nollett}
\address{Department of Physics,
 Enrico Fermi Institute, The University 
of Chicago, Chicago, IL~~60637-1433\\
and\\
Physics Division, Argonne National Laboratory,
         Argonne, Illinois 60439}
\author{R. B. Wiringa}
\address{Physics Division, Argonne National Laboratory,
         Argonne, Illinois 60439}
\author{R. Schiavilla}
\address{Jefferson Lab, Newport News, Virginia 23606 \\
         and \\
         Department of Physics, Old Dominion University,
         Norfolk, Virginia 23529}
\date{\today}
\maketitle

\begin{abstract}
We have computed the cross section for the process
$d(\alpha,\gamma)^6{\rm Li}$ at the low energies relevant for
primordial nucleosynthesis and comparison with laboratory data.  The
final state is a six-body wave function generated by the variational
Monte Carlo method from the Argonne $v_{18}$ and Urbana IX potentials,
including improved treatment of large-particle-separation behavior.
The initial state is built up from the $\alpha$-particle and deuteron
ground-state solutions for these potentials, with phenomenological
descriptions of scattering and cluster distortions.  The dominant $E2$
cross section is in reasonable agreement with the laboratory data.
Including center-of-energy and other small corrections, we obtain an
$E1$ contribution which is larger than the measured contribution at 2
MeV by a factor of seven.  We calculate explicitly the
impulse-approximation $M1$ contribution, which is expected to be very
small, and obtain a result consistent with zero.  We find little
reason to suspect that the cross section is large enough to produce
significant $^6$Li in the big bang.
\end{abstract}

%\vskip 1in
%]
\section{Introduction}

Radiative capture of deuterons on alpha particles is the only process
by which $^6$Li is produced in standard primordial nucleosynthesis
models \cite{nls}.  Because the low-energy cross section for this
process is so small ($< 10^{-2}$ nb), it has long been held that no
measurable amount of $^6$Li can be made in big bang nucleosynthesis
(BBN) without recourse to exotic physics (baryon-inhomogeneous
scenarios, hadronically-evaporating black holes, etc.).  However,
there has been new interest in $^6$Li as a cosmological probe in
recent years, for two reasons.  First, $^6$Li is more sensitive to
destruction in stars than is $^7$Li.  There has consequently been an
attempt to set limits on $^7$Li depletion in halo stars by determining
how much $^6$Li they have destroyed over their lifetimes.  Second, the
sensitivity of searches for $^6$Li has been increasing as a result, so
that there are now two claimed detections in metal-poor halo stars, and 
two more detections in disk stars \cite{hobbs97,smith98,cayrel99,nissen99}.
The explanation of these data in terms of chemical-evolution models
remains an open question \cite{fieldsolive98a,fieldsolive98b,ramaty}.
They are presently at a level exceeding even optimistic estimates of
how much $^6$Li could have been made in standard BBN \cite{nls}, and
the observed $^6$Li is believed to have been created in interactions
between cosmic rays and the interstellar medium.  However, it remains
interesting from the point of view of understanding these observations
to remove as many remaining uncertainties as possible from the
standard scenario.  The main such uncertainty arising in the standard
BBN calculation is the cross section for $d(\alpha,\gamma)^6{\rm Li}$.

At the same time, electroweak processes in light nuclei are
interesting from the point of view of few-body nuclear physics.  The
advent of realistic two- and three-nucleon potentials has enabled
calculation of nuclear wave functions and energy observables of
systems with up to $A=8$ \cite{WPCP00}.  Calculation of electroweak
observables allows both comparison of these wave functions with
experiment ({\it e.g.}, electron scattering form factors \cite{WS98}),
and estimates of effects not observable in the laboratory ({\it e.g.},
weak capture cross sections \cite{SS+98,MSVKR00}).  The radiative
capture of deuterons on alpha particles provides both kinds of
opportunities: there are data on the direct process at 700 keV (all
energies are in center of mass) and above \cite{rghr,mohr}, and there are
indirect data from Coulomb breakup experiments corresponding to
70--400 keV \cite{kiener}.  There is also a need for extrapolation to
lower energies for application to nucleosynthesis.

\begin{figure}
\centerline{\epsfig{file=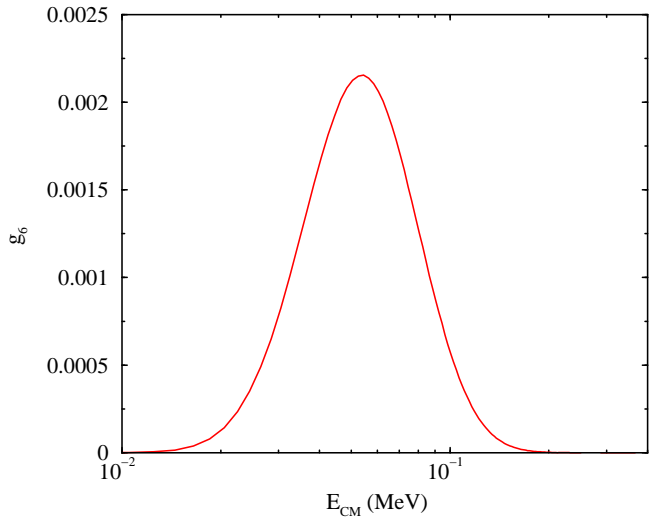,height=8.6cm,angle=270}}
\caption{Sensitivity function of Ref. \protect\cite{nollettburles},
showing the change in $^6$Li BBN yields resulting from small changes
in the \dag cross section over narrow energy bins at the specified
energies.}
\label{fig:sensitivity}
\end{figure}

The $\alpha d$ capture problem is made even more interesting by the
fact that $S$- and $P$-wave captures are strongly inhibited by
quasi-orthogonality between the initial and final states and by an
isospin selection rule, respectively.  As a result, the dominant
process in all experiments performed to date has been electric
quadrupole ($E2$) capture from $D$-wave scattering states.  The small
remaining $E1$ contribution from $P$-wave initial states has been
observed at about 2 MeV, but its magnitude has not been successfully
explained by theoretical treatments; it is generally expected to
contribute half of the cross section at 100 keV.  The $S$-wave capture
induced by $M1$ has been neglected in most calculations because of the
quasi-orthogonality mentioned above, which makes the associated matrix
element identically zero in two-body treatments of the process.  The
energy dependences of the various capture mechanisms ($E2$, $E1$,
$M1$) are such that even $E1$ and $M1$ captures with small amplitudes
may become important at low ($< 200$ keV) energies.  Low-energy
behavior is particularly important for standard BBN: the primordial
$^6$Li yield is only sensitive to the capture cross section between 20
and 200 keV, with the strongest sensitivity at 60 keV.  This is
demonstrated in Fig. \ref{fig:sensitivity}, where we show the
fractional change in \li6 yields resulting from changing the cross
section over narrow bins in energy.  Since it is the energy integral
of this quantity which determines the sensitivity, and it is shown on
a logarithmic axis, the sensitivity function $g_6$ is defined to
include a factor of energy so that relative areas may be accurately
gauged.  (A more extensive discussion of these functions appears in
Ref. \cite{nollettburles}.)  The \li6 yield is directly proportional
to the \dag cross section at the energies indicated by this function,
so that an increase in the cross section by a given factor over the
whole energy range produces an increase of the \li6 yield by the same
factor.

We have carried out a calculation of the alpha-deuteron radiative
capture cross section, treating it as a six-body problem.  The
remainder of this paper describes this calculation, and it is
organized as follows: In Section \ref{sec:wave functions}, we describe
wave functions used to compute the \dag cross section.  In Section
\ref{sec:operators}, we describe the electromagnetic current operators
and the methods used to calculate their matrix elements.  In Section
\ref{sec:results}, we describe our results for the cross section,
thermal reaction rates and nucleosynthesis yields, and in Section
\ref{sec:end}, we summarize our results and discuss their
implications.

\section{wave functions}
\label{sec:wave functions}

\subsection{\h2, \he4, and \li6 ground states}

The wave functions $|\psi_d^{m_d}\rangle$, $|\psi_\alpha\rangle$,
and $|\psi_{{\rm Li}}^{m_6}\rangle$ used in our calculation are
the ground states derived from the Argonne $v_{18}$ two-nucleon~\cite{WSS95} 
and Urbana IX~\cite{PPCW95} three-nucleon potentials, henceforth denoted as 
the AV18/UIX model.  The deuteron wave function is a direct numerical
solution, while the variational Monte Carlo technique described in
Refs.~\cite{W91,APW95,PPCPW97} is used to generate the \he4 and \li6 wave 
functions.  Here $m_d$ and $m_6$ denote spin orientation.

The variational trial functions for light nuclei are constructed from
correlation operators acting on a Jastrow wave function:
\begin{equation}
     |\Psi_T\rangle = \left[ 1 + \sum_{i<j<k} \tilde{U}^{TNI}_{ijk} \right]
              \left[ {\cal S}\prod_{i<j}(1+U_{ij}) \right] |\Psi_J\rangle \ ,
\label{eqn:trial}
\end{equation}
where $U_{ij}$ and $\tilde{U}^{TNI}_{ijk}$ are two- and three-body
correlation operators that include significant spin and isospin dependence
and ${\cal S}$ is a symmetrization operator, needed because the $U_{ij}$ 
do not commute.
For \he4, the Jastrow part takes a relatively simple form:
\begin{equation}
  |\Psi_J\rangle = \prod_{i<j<k \leq 4}f_{ijk}
     \prod_{i<j \leq 4}f(r_{ij}) |\Phi_{\alpha}(0 0 0 0)_{1234}\rangle \ ,
\label{eqn:alpha}
\end{equation}
where $f(r_{ij})$ and $f_{ijk}$ are pair and triplet functions of relative
position only, and $\Phi_{\alpha}(0 0 0 0)$ is a determinant in the
spin-isospin space of the four particles.

For \li6 the Jastrow part has a considerably more complicated structure due
to the need to place the fifth and sixth particles in the p-shell:
\begin{eqnarray}
  |\Psi_J\rangle &=& {\cal A} \left\{
     \prod_{i<j<k \leq 4}f^{sss}_{ijk}
     \prod_{n \leq 4}f^{spp}_{n56}
     \prod_{i<j \leq 4}f_{ss}(r_{ij})
     \prod_{k \leq 4} f_{sp}(r_{k5}) f_{sp}(r_{k6}) \right.  f_{pp}(r_{56})
     \nonumber\\
  && \left.  \sum_{LS} \Big( \beta_{LS} 
     |\Phi_6(LSJMTT_{3})_{1234:56}\rangle \Big) \right\} \ .
\label{eqn:jastrow}
\end{eqnarray}
The ${\cal A}$ is an antisymmetrization operator over all partitions of
the six particles into groups of four plus two.
The central pair and triplet correlations $f_{xy}(r_{ij})$ and
$f^{xyz}_{ijk}$ have been generalized, with the $xyz$ denoting whether 
the particles are in the s- or p-shell.
The wave function $|\Phi_6(LSJMTT_{3})\rangle$ has
orbital angular momentum $L$ and spin $S$ coupled to total angular
momentum $J$, projection $M$, isospin $T$, and charge state $T_{3}$, and
is explicitly written as
\begin{eqnarray}
 &&  |\Phi_{6}(LSJMTT_{3})_{1234:56}\rangle =
     |\Phi_{\alpha}(0 0 0 0)_{1234}
     \phi^{LS}_{p}(R_{\alpha 5}) \phi^{LS}_{p}(R_{\alpha 6})
     \nonumber \\
 &&  \left\{ [Y_{1m_l}(\Omega_{\alpha 5}) Y_{1m_l'}(\Omega_{\alpha 6})]_{LM_L}
     \times [\chi_{5}(\case{1}{2}m_s) \chi_{6}(\case{1}{2}m_s')]_{SM_S}
     \right\}_{JM} \nonumber \\
 &&  \times [\nu_{5}(\case{1}{2}t_3) \nu_{6}(\case{1}{2}t_3')]_{TT_3}\rangle \ .
\end{eqnarray}
Particles 1--4 are placed in the s-shell core with only spin-isospin 
degrees of freedom, while particles 5--6 are 
placed in p-wave orbitals $\phi^{LS}_{p}(R_{\alpha k})$ that are functions of 
the distance between the center of mass of the core and the particle.
Different amplitudes $\beta_{LS}$ are mixed to obtain an optimal wave
function; for the $J^\pi,T=1^+,0$ ground state of $^6$Li the p-shell can have 
$\beta_{01}$, $\beta_{21}$, and $\beta_{10}$ terms.

The two-body correlation operator $U_{ij}$ is defined as:
\begin{equation}
     U_{ij} = \sum_{p=2,6} \left[ \prod_{k\not=i,j}f^p_{ijk}({\bf r}_{ik}
              ,{\bf r}_{jk}) \right] u_p(r_{ij}) O^p_{ij} \ ,
\end{equation}
where the $O^{p=2,6}_{ij}$ = ${\bbox \tau}_i\cdot {\bbox \tau}_j$,
${\bbox \sigma}_i\cdot{\bbox \sigma}_j$,
${\bbox \sigma}_i\cdot{\bbox \sigma}_j {\bbox \tau}_i\cdot {\bbox \tau}_j$,
$S_{ij}$, and $S_{ij}{\bbox \tau}_i\cdot {\bbox \tau}_j$, and
the $f^p_{ijk}$ is an operator-independent three-body correlation.
The six radial functions $f_{ss}(r)$ and $u_{p=2,6}(r)$ are obtained from
two-body Euler-Lagrange equations with variational parameters as discussed
in detail in Ref.~\cite{W91}.  Here we take them to be the same as in
\he4, except that the $u_{p=2,6}(r)$ are forced to go to zero at large 
distance by multiplying in a cutoff factor, 
$\Big[1+{\rm exp}[-R_u/a_u]\Big]/\Big[1+{\rm exp}[(r-R_u)/a_u]\Big]$, 
with $R_u$ and $a_u$ as variational parameters.
The $f_{sp}$ correlation is constructed to be similar to $f_{ss}$ for small 
separations, but goes to a constant of order unity at large distances:
\begin{equation}
f_{sp}(r) = 
\Bigg[ a_{sp} + \frac{b_{sp}}{1+{\rm exp}[(r-R_{sp})/a_{sp}]} \Bigg] f_{ss}(r) 
+ c_{sp}(1-\exp[-(r/d_{sp})^2]) \ , \\
\end{equation}
where $a_{sp}$, $b_{sp}$, etc. are additional variational parameters.
The $f_{pp}(r)$ correlation is given by:
\begin{equation}
f_{pp}(r) = u_d(r)/{[1-3u_2(r)+u_3(r)-3u_4(r)]r} \ ,
\end{equation}
where $u_d(r)$ is the $S$-wave part of the exact deuteron wave function.
Then $U_{ij}$ acting on the pair of nucleons in the p-shell will 
regenerate the correct deuteron $S$-wave and an effective $D$-wave
$w_d(r) = f_{pp}(r)[u_5(r)-3u_6(r)]\sqrt{8}r$ from the tensor components.
While the $D$-wave is not exact, we find that the resulting effective
deuteron energy is $-2.16$ MeV, quite close to the correct value.

These choices for $f_{ss}$, $f_{sp}$, $f_{pp}$, and $u_{p=2,6}$
guarantee that when the two p-shell particles are far from the s-shell core, 
the overall wave function factorizes as:
\begin{equation}
\Psi_T \rightarrow [\phi^{LS}_{p}(R_{\alpha d})]^2 \psi_{\alpha} \psi_d \ ,
\end{equation}
where $\psi_{\alpha}$ is the variational \he4 wave function and $\psi_d$
is (almost) the exact deuteron wave function.

The single-particle functions $\phi^{LS}_{p}(R_{\alpha n})$
describe correlations between the s-shell core and the
p-shell nucleons, and have been taken in previous work~\cite{PPCPW97} to 
be solutions of a radial Schr\"odinger equation for a
Woods-Saxon potential and unit angular momentum, with energy and
Woods-Saxon parameters determined variationally.
It is important for low-energy radiative captures that these functions
reproduce faithfully the large-separation behavior of the
wave function, because the matrix elements receive large contributions
at cluster separations greater than 10 fm.  In fact, below 400 keV,
more than 15\% of the electric quadrupole operator comes from cluster
separations beyond 30 fm.  We have therefore modified the \li6 wave
function for the capture calculation to enforce cluster-like behavior 
when the two p-shell nucleons are both far from the s-shell core.

In general, for light p-shell nuclei with an asymptotic two-cluster
structure, such as $\alpha d$ in \li6 or $\alpha t$ in $^7$Li, we want the 
large separation behavior to be
\begin{equation}
\label{eqn:asymptotic}
[\phi^{LS}_{p}(r\rightarrow\infty)]^n \propto W_{km}(2\gamma r)/r,
\end{equation}
where $W_{km}(2\gamma r)$ is the Whittaker function for bound-state wave 
functions in a Coulomb potential (see below) and $n$ is the number of 
p-shell nucleons.  We achieve this by solving the equation
\begin{equation}
\Bigg[ -\frac{\hbar^2}{2 \mu_{41}} \Bigg( \frac{d^2}{dr^2}
-\frac{\ell(\ell+1)}{r^2} \Bigg) +V(r)+\Lambda(r) \Bigg]
r\phi^{LS}_{p}(r) = 0,
\end{equation}
with $\ell=1$, $\mu_{41}$ the reduced mass of one nucleon against
four, and $V(r)$ a parametrized Woods-Saxon potential plus Coulomb term:
\begin{equation}
V(r) = \frac{V_0}{1+{\rm exp}[(r-R_0)/a_0]} + \frac{2(Z-2)}{n} \frac{e^2}{r} F(r) \ .
\end{equation}
Here $V_0$, $R_0$, and $a_0$ are variational parameters, $(Z-2)/n$ is
the average charge of a p-shell nucleon, and F(r) is a form factor
obtained by folding $\alpha$ and proton charge distributions together.
The $\Lambda(r)$ is a Lagrange multiplier that enforces the asymptotic
behavior at large $r$, but is cut off at small $r$ by means of a
variational parameter $c_0$:
\begin{equation}
\Lambda(r) = \lambda(r) \Big[ 1-{\rm exp}\Big(-(r/c_0)^2\Big) \Big] \ .
\end{equation}
The $\lambda(r)$ is given by
\begin{equation}
\lambda(r) = \frac{\hbar^2}{2 \mu_{41}} \Bigg[\frac{1}{u_L} \frac{d^2 u_L}{dr^2}
- \frac{2}{r^2} \Bigg] - \frac{2(Z-2)}{n} \frac{e^2}{r} \ ,
\end{equation}
where $u_L$ is directly related to the Whittaker function:
\begin{equation}
u_L/r = (W_{km}(2\gamma r)/r)^{1/n} \ .
\end{equation}
Here $\gamma^2 = 2\mu_{4n} B_{4n}/\hbar^2$, with $\mu_{4n}$ and $B_{4n}$
the appropriate two-cluster effective mass and binding energy, 
$k = -2(Z-2) e^2 \mu_{4n}/\hbar^2 \gamma$, and $m = L+\case{1}{2}$.

For \li6, $B_{42} = 1.47$ MeV and $L =0$, 2 corresponding to the
asymptotic $S$- and $D$-waves of the \li6 ground state, or amplitudes
$\beta_{01}$ and $\beta_{21}$ in Eq.(\ref{eqn:jastrow}).  There is no
asymptotic $\alpha d$ cluster corresponding to the $L=1$ amplitude
$\beta_{10}$.  Nevertheless, the energy of the \li6 ground state is
variationally improved $\approx 0.1$ MeV by including such a component
in the wave function, so we set the asymptotic behavior of
$\phi^{10}_{p}(r)$ to a binding energy of 3.70 MeV, which is the
threshold for $^6{\rm Li} \rightarrow\ ^4{\rm He}+p+n$ breakup.

In Refs.~\cite{APW95,PPCPW97} the $f^{sss}_{ijk}$ three-body correlation
of Eq.(\ref{eqn:jastrow}) was a valuable and inexpensive improvement to
the trial function, but no $f^{ssp}_{ijk}$ or $f^{spp}_{ijk}$ correlations
could be found that were of any benefit.  However, in the present work
we find the correlation
\begin{equation}
f^{spp}_{n56} = 1 + q_1 [f_{ss}(r_{56})/f_{pp}(r_{56}) - 1] 
                        {\rm exp}[-q_2(r_{n5}+r_{n6})] \ ,
\end{equation}
with $q_{1,2}$ as variational parameters,
to be very useful.  It effectively alters the central pair correlation between 
the two p-shell nucleons from their asymptotic, deuteron-like form to be more 
like the pair correlations within the s-shell when the two particles are close 
to the core.  This correlation improves the binding energy by $\approx 0.5$ MeV.

In Refs.~\cite{WPCP00,PPCPW97} we reported energies for the trial function
$\Psi_T$ of Eq.(\ref{eqn:trial}), and for a more sophisticated variational 
wave function, $\Psi_V$, which adds two-body spin-orbit and three-body spin- 
and isospin-dependent correlation operators.  The $\Psi_V$ gives improved 
binding compared to $\Psi_T$ in both \he4 and \li6, but is significantly 
more expensive to construct because of the numerical derivatives required 
for the spin-orbit correlations.  In the case of an energy calculation
the derivatives are also needed for the evaluation of $L$-dependent terms
in AV18, so the cost is only a factor of two in computation.  However for
the evaluation of other expectation values the relative cost increase is 
$\approx 6A$.
Thus in the present work we choose to use $\Psi_T$ for non-energy evaluations;
this proved quite adequate in our studies of \li6 form factors~\cite{WS98}.

The variational Monte Carlo (VMC) energies and point proton rms radii
obtained with $\Psi_T$ and $\Psi_V$ are shown in Table~\ref{tab:energy}
along with the results of essentially exact Green's function Monte Carlo
(GFMC) calculations and the experimental values.
We note that the underbinding of \li6 in the GFMC calculation is the fault
of the AV18/UIX model and not the many-body method; it can be corrected by 
the introduction of more sophisticated three-nucleon potentials~\cite{PPRWC00}.

\begin{table}
\caption{Calculated VMC, GFMC, and experimental energies and point proton
rms radii of $^2$H, $^4$He and $^6$Li. Numbers in parentheses are Monte
Carlo statistical errors, or experimental uncertainties.}
\begin{tabular}{ccdddd}
Nucleus & Observable & \multicolumn{2}{c}{VMC} & GFMC & Experiment \\
        &            & $\Psi_T$  & $\Psi_V$    &      &            \\
\hline
$^2$H  & E &  --2.2246    &           &            &  --2.2246 \\
       & $\langle r^2_p \rangle^{1/2}$
           &    1.967    &            &            &    1.953(3) \\
$^4$He & E & --26.89(3)  & --27.78(3) & --28.34(4) & --28.30 \\
       & $\langle r^2_p \rangle^{1/2}$
           &    1.48(1)  &    1.47(1) &    1.45(1) &    1.48(1) \\
$^6$Li & E & --27.19(5)  & --28.21(4) & --31.15(11)& --31.99 \\
       & $\langle r^2_p \rangle^{1/2}$
           &    2.46(2)  &    2.46(2) &    2.57(1) &    2.43(4) \\
\end{tabular}
\label{tab:energy}
\end{table}

\begin{figure}
\centerline{\epsfig{file=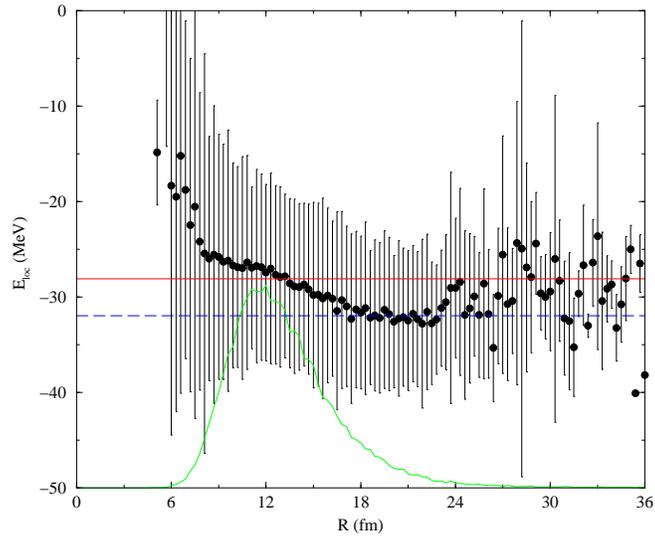,height=8.6cm,angle=270}}
\caption{Monte Carlo local energy of the \li6 variational ground state
binned as a function of the sum of the magnitudes of the particle coordinates, 
$R$.  The horizontal solid line is the average variational energy and the
horizontal dashed line indicates the experimental binding.
The curve superimposed on the bottom is a histogram of the location
of the 50 000 samples of $\Psi_V$ used in the calculation.}
\label{fig:energy}
\end{figure}

The present variational trial functions with the imposed proper asymptotic
behavior in fact give slightly better energies than the older shell-model-like
correlations of Refs.~\cite{WPCP00,PPCPW97}.
Unfortunately, because the variational \li6 energy is not below that of 
separated alpha and deuteron clusters, it is possible to lower the energy
significantly by making the wave function more and more diffuse.
Therefore, the variational parameters were constrained to give a reasonable
rms radius, as seen in Table~\ref{tab:energy}.
Although the variational \li6 wave function remains
unbound relative to alpha-deuteron breakup, it has the correct energy
at large particle separations, as shown in Fig.~\ref{fig:energy}.  There
we show the local energy, $E({\bf R})$, which is the energy for a given 
spatial configuration ${\bf R}=({\bf r}_1,\dots,{\bf r}_6)$,
binned with its Monte Carlo statistical 
variance as a function of the sum $R = \sum_i |{\bf r}_i|$ of the particle 
distances from the center of mass.
The average energy is shown by the solid line, while the dashed line
gives the experimental binding.
At the bottom of the figure is a histogram of the number of samples to
show their distribution with $R$.
We observe that the energy in compact configurations is clearly too small, 
but for $R > 18$ fm the energies scatter about the experimental binding.

\begin{figure}
\centerline{\epsfig{file=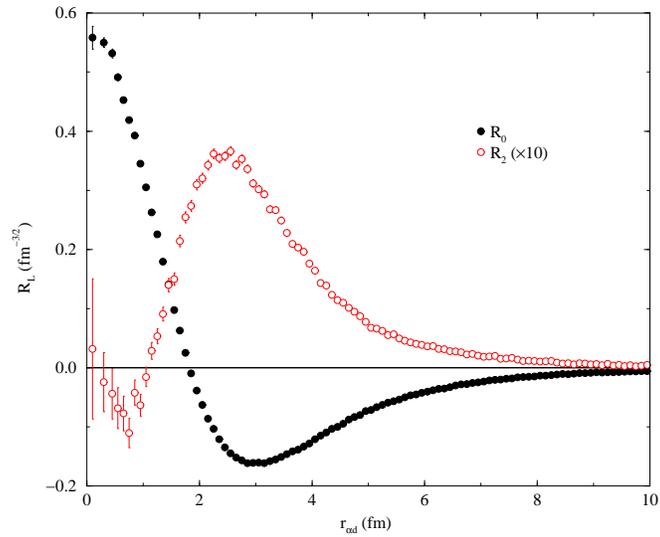,height=8.6cm,angle=270}}
\caption{Monte Carlo samples of the radial two-cluster $\alpha d$ 
distribution functions in \li6.}
\label{fig:rls}
\end{figure}

The asymptotic two-cluster behavior of our six-body wave function can be
studied by computing the two-cluster $\alpha d$ distribution function,
$\langle{\cal A}\psi_{\alpha}\psi_d^{m_d},{\bf r}_{\alpha d}\mid\psi_{\rm Li}^{m_6}\rangle$,
described in Ref.~\cite{FPPWSA96}.  It can be expressed in terms of 
Clebsch-Gordan factors, $Y_{LM}(\Omega_{\alpha d})$, and two radial functions 
$R_L(r_{\alpha d})$ for the $S$- and $D$-waves, which are plotted in 
Fig.~\ref{fig:rls}.
The $R_L$ can be used to extract the asymptotic normalization constants
$C_0$ and $C_2$, by taking their ratio with the corresponding Whittaker
functions, and the asymptotic $D/S$ ratio $\eta$.  
We find that $C_0$ does not become asymptotic until 
$r_{\alpha d} \agt 8$ fm, where we obtain $C_0 = 2.28 \pm 0.02$.
However, $\eta$ is already asymptotic for $r_{\alpha d} \agt 4$ fm, with the 
value $-0.026 \pm 0.001$.

This value of $C_0$ is at the lower end of the range of values 
(2.3--2.9) used in a number of other studies~\cite{ryzhikh,mukh,GK99}.
However, the value of $\eta$ is consistent with the recent analysis of
\li6 + \he4 elastic scattering~\cite{GK99}, which obtains $-0.025 \pm
0.006 \pm 0.010$.  This is a big improvement over our earlier value of
$-0.07 \pm 0.02$ reported in Ref.~\cite{FPPWSA96}, which was obtained
with the shell-model-like wave function.  However, our quadrupole
moment $Q = -0.71 \pm 0.08$ fm$^2$ remains an order of magnitude too
large compared to the experimental value of $-0.08$ fm$^2$.  The
integrated $D^{\alpha d}_2$ parameter value is $-0.27 \pm 0.01$
fm$^2$.

The spectroscopic factor is 0.86, in good agreement with the Robertson 
value of $0.85 \pm 0.04$~\cite{rghr}.  The spectroscopic factor is not very
sensitive to the asymptotic behavior of the wave function, and differs little 
from our earlier value of 0.84 obtained with the shell-model-like wave function.
We note that our shell-model-like wave functions for $^7$Li and $^6$He
provided an excellent prediction for the spectroscopic factors observed in
$^7{\rm Li}(e,e^\prime p)$ scattering~\cite{LWW99}.

\subsection{${\bf \alpha}$d initial state}

The initial alpha-deuteron scattering wave
function $|\psi_{\alpha d};LSJM\rangle$ having
orbital angular momentum $L$ and channel spin $S$ (=1)
coupled to total $JM$, is expressed as
\begin{equation}
\label{eqn:scatstate}
|\psi_{\alpha d}; LSJM \rangle = 
{\cal A}\left\{\phi_{\alpha d}(r_{\alpha d})Y_{LM_L}({\bf\hat{r}}_{\alpha d})
\prod_{ij}G_{ij}|\psi_\alpha\psi_d^{m_d}\rangle\right\}_{LSJM},
\end{equation}
where curly braces indicate angular momentum coupling, $\cal{A}$
antisymmetrizes between clusters, and ${\bf r}_{\alpha d}$ denotes
the separation between the centers of mass of the $\alpha$
and $d$ clusters.

Our scattering wave function is built from the alpha and deuteron
ground-state wave fuctions, $\psi_\alpha$ and $\psi_d^{m_d}$ (where
$m_d$ denotes the deuteron spin projection), from a nucleon pair
correlation operator $G_{ij}$, and from an intercluster scalar
correlation $\phi_{\alpha d} (r_{\alpha d})$.  $G_{ij}$ is the
identity operator if particles $i$ and $j$ belong to the same cluster
(alpha or deuteron).  Otherwise, it is a pair correlation operator
derived to describe nuclear matter~\cite{LP81b}, and it contains both
central and non-central terms.  It takes into account, in an
approximate way, the distortions introduced in each cluster by the
interactions with the particles in the other cluster, and becomes the
identity operator as particle separations become large.

The intercluster correlation $\phi_{\alpha d}(r_{\alpha d})$ is a
two-body radial wave function describing alpha-deuteron scattering as
the scattering of two point particles.  It is generated from a
Schr\"odinger equation, with the potential chosen to reproduce the
results of phase-shift analyses of alpha-deuteron scattering data in
the two-body model.  Several such potentials exist in the literature,
and we saw no need to re-fit the data.  The potentials which we have
applied to this study are the potentials of Kukulin and Pomerantsev,
Refs. \cite{kp1,kp2} (hereafter, ``KP I'' and ``KP II'',
respectively), and the potential of Langanke, Ref. \cite{langanke1}.
Because the potential of Langanke was produced for the slightly
different problem of describing the $\alpha d$ system in the
resonating group method, it required modification to match our
purposes.  This consisted of a small adjustment to the Woods-Saxon
radius to reproduce the energy of the $3^+$ resonance derived from
electron scattering, 711 keV.  The most important difference between
these potentials, which give very similar cross sections in our
calculations, is that the KP potentials were fitted to the phase
shifts in the odd and even partial waves separately, to produce the
expected parity-dependent potential.  The work by Langanke was
concerned only with even partial waves, and did not fit the odd
partial waves, although it provides a fair qualitative description of
the $P$-wave phase shifts at low energy.

The potentials are very similar to each other in form.  All of the
potentials used contain both central and spin-orbit terms, and the
spin-orbit force in the even partial waves is well-constrained by the
spacing of the $D$-wave scattering resonances.  While a tensor
component could in principle be derived to reproduce the ground-state
quadrupole moment (assuming total alpha-deuteron cluster parentage of
the ground state), its effect on the scattering data is too small to
support deriving it from the scattering phase shifts.  (A more
ambitious approach to derive a tensor interaction would involve
computing energies at fixed separation between the $\alpha$ and $d$
clusters and varying the deuteron orientation; however, we have not
pursued such an approach here.)  Coupling to other ({\it e.g.,}
cluster breakup) channels has also been neglected; this is
well-justified below the $^5{\rm He} + p$ and $^5{\rm Li}+n$
thresholds at 3.12 and 4.20 MeV, respectively.  We note that although
the KP potentials and the Langanke potential all reproduce the
scattering phase shifts very well, and they produce very similar
ground-state wave functions at small cluster separation, the
ground-state binding of the KP II potential is too large by almost a
factor of two.  All of the potentials we used to generate
$\phi_{\alpha d}$ are deep potentials with a forbidden zero-node
state.  They produce the expected one-node structure of the
ground-state alpha-deuteron wave function required by the Pauli
principle, as illustrated by our $R_L$ functions in Fig.\ref{fig:rls},
and enforce (by orthogonality to the forbidden state) the most
important consequences of the Pauli principle for $\phi_{\alpha d}$.

A complication arises from the phenomenological treatment of the
initial state.  In reality, the ground-state and $S$-wave scattering-state 
wave functions are orthogonal, because they are different
eigenstates of the same ($J^\pi=1^+$) potential.  Because our
calculation does not generate them explicitly as such, our ground
state and $S$-wave scattering states are not necessarily orthogonal.
For purposes of our capture calculation, this has the effect of
generating spurious contributions to the $M1$ transition operator.
For $S$-wave captures, we have modified the initial state to ensure
orthogonality with the ground state.  This is done by subtracting from
the initial state, Eq.(\ref{eqn:scatstate}), its projection onto a
complete set of ground states,
\begin{equation}
|\psi_{\alpha d};LSJM \rangle^\prime = 
| \psi_{\alpha d};LSJM \rangle 
- \sum_{m_6} |\psi_{\rm Li}^{m_6}\rangle \langle \psi_{\rm Li}^{m_6}
|\psi_{\alpha d};LSJM \rangle \ ,
\end{equation}
and computing $M1$ transitions from the adjusted state.  This
procedure reduced the $M1$ contribution to our computed cross sections
by more than a factor of 10, roughly to the level of Monte Carlo
sampling noise.  Further evidence that the remaining contribution is
spurious is its strong dependence on the phenomenological radial
correlation in the scattering state, discussed below.

\section{Cross Section and Transition Operators}
\label{sec:operators}

The cross section for $\alpha$$d$ radiative capture at c.m. energy
$E_{c.m.}$ can be written as~\cite{enchilada}

\begin{equation}
\sigma(E_{c.m.})=\frac{8\pi}{3} \frac{\alpha}{v_{\rm rel}}
\frac{q}{1+q/m_{\rm Li}} \sum_{LSJ\ell}\Bigg[ \left|E^{LSJ}_\ell(q)\right|^2
+\left|M^{LSJ}_\ell(q)\right|^2 \Bigg] \ ,
\end{equation}
where $\alpha$ is the fine structure constant ($\alpha=e^2$; we take
$\hbar=c=1$ in this Section), $v_{\rm rel}$ is the $\alpha$$d$
relative velocity, and $E^{LSJ}_\ell(q)$ and $M^{LSJ}_\ell(q)$ are the
reduced matrix elements (RMEs) of the electric and magnetic multipole
operators connecting the $\alpha$$d$ scattering state in channel $LSJ$
to the $^6$Li ground state having $J^\pi,T=1^+$,0.  The c.m. energy
of the emitted photon is given by

\begin{eqnarray}
q&=&m_{\rm Li}\Bigg[ -1 + \sqrt{1+\frac{2}{m_{\rm Li}}
(m_d+m_\alpha-m_{\rm Li}+E_{c.m.})} \Bigg] \nonumber \\
&\simeq&m_d+m_\alpha-m_{\rm Li}+E_{c.m.} \ ,
\end{eqnarray}
where $m_d$, $m_\alpha$, and $m_{\rm Li}$ are the rest masses
of deuteron, $^4$He, and $^6$Li, respectively.  The astrophysical
$S$-factor is then related to the cross section via

\begin{equation}
S(E_{c.m.}) = E_{c.m.}\, \sigma(E_{c.m.})\,
{\rm exp}( 4\, \pi \, \alpha/v_{\rm rel}) \ .
\end{equation}
The $M1$, $M2$, $E1$, $E2$, and $E3$ transitions involving $\alpha$$d$
scattering states with relative orbital angular momentum up to $L=2$
have been retained in the evaluation of the cross section.  Initial
states with $L=3$ and $L=4$ were examined in the early phases of this
work, but proved to be very small, and were not retained in the final
calculation.

Since the energies of interest in the present study are below 4 MeV,
and consequently $qR_{\rm Li} \leq 0.05$ ($R_{\rm Li}$ is the $^6$Li point
rms radius), the long-wavelength aproximation (LWA) of the multipole
operators would naively be expected to be adequate to compute the
associated RMEs.  While the LWA is indeed sufficient for $E2$
transitions, which dominate the cross section at energies $E_{c.m.} >
400$ keV, it is however inaccurate to compute the RMEs of $M1$ and,
particularly, $E1$ transitions in the LWA since these are suppressed.
(This fact has already been alluded to in the introduction).  It is
therefore necessary to include higher order terms, beyond the leading
one which normally is taken to define the LWA, in the expansion in
powers of $q$, so-called ``retardation terms''.  It is useful to
review how these corrections arise for $E1$ transitions.  We should
point out that some of these same issues have recently been discussed
in Ref.~\cite{Viv00}, in the context of a calculation of $E1$ strength
in $d$$p$ radiative capture at energies below 100 keV.

\subsection{The $E_1$ operators}

In principle, the $E1$ RMEs are obtained from

\begin{equation}
\label{eq:e1me}
E_1^{LSJ}(q) = {\sqrt{3} \over \langle JM, 1\lambda | 1m_6\rangle }
\langle \Psi_{\rm Li}^{m_6} |E_{1\lambda}(q)|
\psi_{\alpha d};LSJM\rangle \ ,
\end{equation}

\begin{equation}
\label{eq:e1}
  E_{1\lambda}(q) = {1\over q}
  \int d{\bf x}\> {\bf j}({\bf x}) \cdot \nabla \times j_1(qx)
  {\bf Y}_{1\lambda}^{11}(\hat{\bf x})  \ ,
\end{equation}
where  ${\bf j}({\bf x})$ is the nuclear current density
operator, $j_1(qx)$ is the spherical Bessel function of order one,
and ${\bf Y}_{1\lambda}^{11}(\hat{\bf x})$ are vector spherical
harmonic functions.  For ease of presentation,
the factor $\sqrt{3}/ \langle JM, 1\lambda | 1m_6\rangle$
occurring in the definition of the RMEs will be suppressed in
the following discussion.
By expanding the Bessel function in powers of $q$ and
after standard manipulations~\cite{WAL95}, the $E_1$ operator
can be expressed as

\begin{equation}
\label{eq:lw}
  E_{1\lambda}(q) \simeq  E_{1\lambda}(q;{\rm LWA1})
                     + E_{1\lambda}(q;{\rm LWA2})
                     + E_{1\lambda}(q;{\rm LWA3})\>\>\> ,
\end{equation}
where

\begin{equation}
\label{eq:lw1}
  E_{1\lambda}(q;{\rm LWA1}) =
  -{\sqrt{2} \over 3} \left[ H \, , \,
  \int d{\bf x}\> x\,Y_{1\lambda}(\hat{\bf x})
\>\rho ({\bf x}) \right]  \>\>\> ,
\end{equation}

\begin{equation}
\label{eq:lw2}
  E_{1\lambda}(q;{\rm LWA2}) = { {\rm i}\, q^2 \over 3 \sqrt{2} }
  \int d{\bf x}\, x \, Y_{1\lambda}(\hat{\bf x})\, {\bf x}
\cdot {\bf j}({\bf x}) \>\>\>,
\end{equation}

\begin{equation}
\label{eq:lw3}
  E_{1\lambda}(q;{\rm LWA3}) =
  {\sqrt{2}\, q^2 \over 15} \left[ H \, , \,
  \int d{\bf x}\> x^3 \, Y_{1\lambda}(\hat{\bf x})
\> \rho({\bf x}) \right] \>\>\>.
\end{equation}
Here the continuity equation has been used
to relate $\nabla \cdot {\bf j}({\bf x})$
occurring in $E_{1\lambda}(q;{\rm LWA1})$ and $E_{1\lambda}(q;{\rm LWA3})$ to
the commutator $-{\rm i} [ H \, ,\, \rho({\bf x})]$,
where $\rho({\bf x})$ is the charge density operator.  Evaluating
the RMEs of these operators leads to

\begin{equation}
\label{eq:lwme}
E_1^{LSJ}(q) \simeq E_1^{LSJ}(q;{\rm LWA1})+E_1^{LSJ}(q;{\rm LWA2}) 
+E_1^{LSJ}(q;{\rm LWA3}) \ ,
\end{equation}
with

\begin{equation}
\label{eq:lw1me}
E_1^{LSJ}(q;{\rm LWA1}) =
{\sqrt{2}\, q \over 3} \langle \Psi_{\rm Li}^{m_6} |
\int d{\bf x}\, x Y_{1\lambda}(\hat{\bf x}) \rho({\bf x})
|\psi_{\alpha d};LSJM\rangle \ ,
\end{equation}

\begin{equation}
\label{eq:lw2me}
E_1^{LSJ}(q;{\rm LWA2}) =
{ {\rm i}\, q^2 \over 3 \sqrt{2} } \langle \Psi_{\rm Li}^{m_6} |
\int d{\bf x}\, x Y_{1\lambda}(\hat{\bf x}) {\bf x} \cdot {\bf j}({\bf x})
|\psi_{\alpha d};LSJM\rangle \ ,
\end{equation}

\begin{equation}
\label{eq:lw3me}
E_1^{LSJ}(q;{\rm LWA3}) =-
{\sqrt{2}\, q^3 \over 15} \langle \Psi_{\rm Li}^{m_6} |
\int d{\bf x}\, x^3 Y_{1\lambda}(\hat{\bf x}) \rho({\bf x})
|\psi_{\alpha d};LSJM\rangle \ ,
\end{equation}
where terms up to order $q^3$ have been retained, and it has been
assumed that the initial and final VMC wave functions are exact
eigenfunctions of the Hamiltonian -- which, incidentally, is clearly
not the case, see the previous Section -- so that $[H\, ,\, \rho({\bf
x})] \rightarrow - q \, \rho({\bf x})$ in the matrix element, ignoring
the kinetic energy of the recoiling $^6$Li.

The nuclear electromagnetic charge and current density operators
have one- and two-body components.  

\begin{eqnarray}
\rho({\bf x})&=& \sum_i \rho^{(1)}_i({\bf x})
             +\sum_{i<j} \rho^{(2)}_{ij}({\bf x}) \>\>; \label{eq1}\\
{\bf j}({\bf x})&=& \sum_i {\bf j}^{(1)}_i({\bf x})
             +\sum_{i<j} {\bf j}^{(2)}_{ij}({\bf x}) \label{eq2} \>\>.
\end{eqnarray}
The one-body terms $\rho^{(1)}_i$ and ${\bf j}^{(1)}_i$ have the
standard expressions~\cite{WAL95} obtained from a non-relativistic
reduction of the covariant single-nucleon current.  The charge density
is written as
\begin{equation}
\rho^{(1)}_i({\bf x})= \rho^{(1)}_{i,{\rm NR}}({\bf x})+
                       \rho^{(1)}_{i,{\rm RC}}({\bf x}) \>\>, \label{eq6}
\end{equation}
with
\begin{equation}
\rho^{(1)}_{i,{\rm NR}}({\bf x})= \epsilon_i \>
  \delta({\bf x}-{\bf r}_i) \label{eq7} \>\>,
\end{equation}
\begin{equation}
\rho^{(1)}_{i,{\rm RC}}({\bf x})=
 {\frac {1}{8m^2}} \left ( 2\, \mu_i-\epsilon_i \right )
\nabla_x \cdot \Bigg [ \bbox{\sigma}_i \times {\bf p}_i \, ,\, 
\delta({\bf x}-{\bf r}_i) \Bigg ]_+ \>\>,
\label{eq8}
\end{equation}
where the Darwin-Foldy relativistic correction
has been neglected in the expression
for $\rho^{(1)}_{i,{\rm RC}}$, since it gives no contribution
to Eq.~(\ref{eq:lw1me}) (see below).
Hereafter, $[ \cdots \, ,\, \cdots ]_+$ denotes the anticommutator.
The one-body current density is expressed as

\begin{equation}
{\bf j}^{(1)}_i({\bf x})={\frac {1} {2m}} \epsilon_i \>
\Biggl[ {\bf p}_i\>,\> \delta( {\bf x}-{\bf r}_i) \Biggr ]_+
-{\frac {1} {2m}} \mu_i \>
  \bbox{\sigma}_i \times \nabla_x \delta({\bf x}-{\bf r}_i) \ .
\label{eq9}
\end{equation}
The following definitions have been introduced:

\begin{eqnarray}
\epsilon_i &\equiv& \frac{ 1 + \tau_{i,z}}{2} \>\>,
\label{eq8a}\\
\mu_i &\equiv& \frac{\mu^S + \mu^V\tau_{i,z}}{2}
\label{eq9a} \>\>,
\end{eqnarray}
where $\mu^S$ ($\mu^S$=0.88 n.m.) and $\mu^V$ ($\mu^V$=4.706 n.m.) are
the isoscalar and isovector combinations of the proton and neutron
magnetic moments.

The two-body charge and current density operators have
pion range terms and additional, short-range terms due to heavy
meson exchanges~\cite{enchilada}.  However, the pion-exchange current
density is an isovector operator, and its matrix element vanishes
identically in the radiative $\alpha$$d$ capture, while the heavy-meson
exchange (charge and current) operators are
expected to give negligible contributions.
Hence, in the present study we only retain the charge density
operator associated with pion exchange.  It is explicitly given by

\begin{equation}
\rho^{(2)}_{ij,\pi}({\bf x})=-\frac{f^2_\pi}{2\,m\, m_\pi^2}
\bbox{\tau}_i \cdot \bbox{\tau}_j\, I_\pi^\prime(r_{ij}) \,
\Bigg[\bbox{\sigma}_j \cdot \hat{\bf r}_{ij}
\bbox{\sigma}_i\cdot \nabla_x \,\delta({\bf x}-{\bf r}_i)\,
+i\rightleftharpoons j \Bigg] \ ,
\end{equation}
where $m_\pi$ is the charged pion mass,
$f_\pi$ the pion-nucleon coupling constant ($f_\pi^2/4\pi=0.075$),
${\bf r}_{ij}={\bf r}_i-{\bf r}_j$ is the
separation of nucleons $i$ and $j$, and the function
$I_\pi(r)$ is defined as

\begin{equation}
I_\pi(r)=\frac{1}{4\pi r}\left[{\rm e}^{-m_\pi r}-{\rm e}^{-\Lambda_\pi r}
-\frac{1}{2}\left(1-\frac{m_\pi^2}{\Lambda_\pi^2}\right)
\Lambda_\pi r {\rm e}^{-\Lambda_\pi r}\right] \ .
\end{equation}
The prime denotes differentiation with respect to its argument.  For
the parameter $\Lambda_\pi$, we have taken the value 1.05 GeV.

We now consider, in turn, the various contributions
LWA1, LWA2, and LWA3 to the $E_1$ RME.
Insertion of the NR, RC, and $\pi$-exchange charge
density operators in Eq.~(\ref{eq:lw1me}) allows us to
write correspondingly 

\begin{equation}
\label{eq:lwcb}
  E_1^{LSJ}(q;{\rm LWA1}) \simeq  E_1^{LSJ}(q;{\rm LWAc})
                              + E_1^{LSJ}(q;{\rm LWAb})
                              + E_1^{LSJ}(q;{\rm LWA}\pi)
\end{equation}
where

\begin{equation}
\label{eq:LWAc}
E_1^{LSJ}(q;{\rm LWAc})=
{\sqrt{2}\, q \over 3} \langle \Psi_{\rm Li}^{m_6} |
\sum_i \epsilon_i\, r_i \, Y_{1\lambda}(\hat{\bf r}_i)
|\psi_{\alpha d};LSJM\rangle \ ,
\end{equation}

\begin{equation}
\label{eq:lwb}
  E_1^{LSJ}(q;{\rm LWAb}) =
  {\sqrt{2}\, q \over 3} \sqrt{\frac{3}{4\pi}}\langle \Psi_{\rm Li}^{m_6} |
 \sum_i - { 2 \mu_i-\epsilon_i \over 4 m^2 }
(\bbox{\sigma}_i \times {\bf p}_i)_\lambda
  |\psi_{\alpha d};LSJM\rangle \ ,
\end{equation}

\begin{eqnarray}
\label{eq:lwp}
  E_1^{LSJ}(q;{\rm LWA}\pi) =
  {\sqrt{2}\, q \over 3} \sqrt{\frac{3}{4\pi}}\langle \Psi_{\rm Li}^{m_6} |
 \sum_{i<j}&& \frac{f_\pi^2}{2\,m\, m_\pi^2}
\bbox{\tau}_i \cdot \bbox{\tau}_j\, I_\pi^\prime(r_{ij}) \,
\Bigg[\sigma_{i,\lambda} \, \bbox{\sigma}_j \cdot \hat{\bf r}_{ij} \nonumber \\
&&+i \rightleftharpoons j\Biggr] |\psi_{\alpha d};LSJM\rangle \ ,
\end{eqnarray}
where $(\sigma \times {\bf p})_\lambda$ and $\sigma_{i,\lambda}$
denote spherical components.

Next, for the terms of order $q^2$ and $q^3$, Eqs.~(\ref{eq:lw2me}) and
(\ref{eq:lw3me}), we find:

\begin{eqnarray}
\label{eq:lw22}
E_1^{LSJ}(q;{\rm LWA2}) =
{ {\rm i}\, q^2 \over 3 \sqrt{2} } \langle \Psi_{\rm Li}^{m_6} |
\sum_i&&\frac{\epsilon_i}{2\,m} \Bigg [ {\bf p}_i \cdot \, ,\, 
{\bf r}_i \,r_i\, Y_{1\lambda}(\hat{\bf r}_i) \Bigg ]_+ \nonumber \\
- &&\frac{\mu_i}{2\,m} \sqrt{\frac{3}{4\pi}}
( {\bf r}_i \times \bbox{\sigma}_i )_\lambda 
|\psi_{\alpha d};LSJM\rangle \ ,
\end{eqnarray}

\begin{equation}
\label{eq:lw33}
E_1^{LSJ}(q;{\rm LWA3}) =-
{\sqrt{2}\, q^3 \over 15} \langle \Psi_{\rm Li}^{m_6} |
\sum_i \epsilon_i \, r_i^3 \, Y_{1\lambda}(\hat{\bf r}_i)
|\psi_{\alpha d};LSJM\rangle \ ,
\end{equation}
where in the last equation the contributions from relativistic
corrections and pion-exchange have been neglected. 

The leading contribution $E_1^{LSJ}(q;{\rm LWAc})$ to the $E_1$ RME
vanishes in $\alpha$$d$ capture, since the initial and final
states have zero isospin, and the corresponding wave functions do
not depend on the c.m. position:

\begin{equation}
\label{lwaccancel}
E_1^{LSJ}({\rm LWAc})\propto
\langle \Psi_{\rm Li}^{m_6} |
\sum_i {\bf r}_{i,\lambda} |\psi_{\alpha d};LSJM\rangle = 0\ ,
\end{equation}
where the coordinates ${\bf r}_i$ are defined relative to that of the
c.m..  

There are also more subtle corrections, which can potentially lead to
additional, (relatively) significant $E_1$ contributions.  One such
correction is the possibility of isospin admixtures with $T > 0$ in
the $^6$Li and $\alpha$$d$ states, originating from the
isospin-symmetry-breaking interactions present in the AV18/UIX model.
These $T>0$ components are ignored in the VMC wave functions used
here.

An additional correction -- which we take into account -- arises from
a relativistic correction to the center of mass, and makes the LWAc
contribution nonzero.  This correction arises because
translation-invariant wave functions require not center-of-mass but
center-of-energy coordinates.  Such coordinates constitute the
relativistic analog of the center-of-mass coordinates appropriate to
nonrelativistic problems.  The distinction is usually not crucial at
low energies, but it matters here because the LWAc term would
otherwise be zero.  Since the coordinates in which we compute wave
functions are not center-of-energy coordinates, the dipole moment

\begin{equation}
{\bf d} = \sum_i \epsilon_i {\bf r}_i,
\end{equation}
which enters Eq. (\ref{lwaccancel}), should be replaced by the
expression
\begin{eqnarray}
{\bf d} & = & \sum_i \epsilon_i ({\bf r}_i-{\bf r}_{CE})\\
& =  & -\sum_i \epsilon_i {\bf r}_{CE} \ ,
\end{eqnarray}
where
\begin{equation}
{\bf r}_{CE}=\frac{\sum_i E_i {\bf r}_i}{\sum_i E_i} \ ,
\end{equation}
and
\begin{equation}
E_i=m+\frac{p_i^2}{2m}+\frac{1}{2}\sum_{j\neq i}v_{ij}
+\frac{1}{3}\sum_{j\neq k\neq i}V_{ijk} \ .
\end{equation}
Here, $v_{ij}$ and $V_{ijk}$ are the two- and three-nucleon
potentials.  We have thus replaced the particle masses in the
expression for center of mass by their total energies (rest energy
plus kinetic and potential energies).  Since the nucleon masses
contribute most of the nucleons' relativistic energies, the center of
mass is only slightly different from the center of energy.  (The rms
radius of the center of energy for our $^6$Li variational ground state
is $(9.1\pm 2.7)\times 10^{-3}$ fm.)  We note that this correction has
been taken into account in the past (in two-body models) by putting
the measured alpha and deuteron masses into calculations, rather than
$4m$ and $2m$ \cite{rghr,ryzhikh,mukh,burkova,typel}.  These differ by
the mass defects of the alpha particle and deuteron.

The $E1$ cross sections arising from the center-of-energy correction
are larger than the $E1$ cross sections measured in Ref. \cite{rghr}
by about a factor of six.  There is a significant ($\simeq 30\%$) LWAb
spin-orbit contribution which becomes more important with increasing
energy; the next-largest contributions are from the LWA3 and LWA2
operators.  All of these terms except LWAb are too small in our
calculation to account for the observed transition by more than a
factor of ten, but LWA3 is responsible for about 20\% of the total $S$
factor at zero energy.  We have also included the contribution
labelled LWA$\pi$ in the present study, and we find it to be much
smaller than LWAb.  Note that although the LWAb term is large enough
to account for the magnitude of the $E1$ data alone, it is of the
wrong sign to account for the asymmetry from which the $E1$ strength
is extracted.

\subsection{The Monte Carlo Calculation}

We calculated matrix elements by modifying the variational Monte Carlo
code described in Ref. \cite{PPCPW97} to perform Monte Carlo
integrations of matrix elements between the scattering and ground
states described above.  The Monte Carlo algorithm used was the
Metropolis algorithm, and we tried several weighting functions, all
based on the \li6 ground state.  The final calculation used the
weighting function

\begin{equation}
W({\bf R})=\sqrt{\langle \psi_{\rm Li}^{m_6}({\bf R})
| \psi_{\rm Li}^{m_6}({\bf R}) \rangle}
\end{equation}
(where the bra-ket product here indicates summation over spin and
isospin degrees of freedom for a given spatial configuration ${\bf
R}=({\bf r}_1,{\bf r}_2,\ldots,{\bf r}_6)$ of the particles; in all
other cases, it also indicates integration over space coordinates).
This weighting function was chosen to reflect approximately the
expected behavior of the matrix element, and to obtain good sampling
both at small cluster separation and out to 40 fm cluster separation.
In the low-energy calculation, 3\% of the $E1$ matrix element and 15\%
of the $E2$ matrix element come from the region beyond 30 fm cluster
separation.  

Because all of the energy dependence is contained in the relative
wave function $\phi_{\alpha d}(r_{\alpha d})$ and the transition operators,
the matrix element for a given scattering partial wave can be
re-written as
\begin{equation}
\label{eqn:integrand}
T^{LSJ}_\ell(q)=\int_0^\infty dx\, x^2 \,
\phi_{\alpha d}(x)\langle \psi_{\rm Li}^{m_6}|
T_{\ell \lambda}(q)
{\cal A}\left\{\delta(x-r_{\alpha d} )
Y_L^{M_L}({\bf\hat{r}}_{\alpha d}))
\prod_{ij}G_{ij}|\psi_\alpha\psi_d^{m_d}\rangle\right\}_{LSJM},
\end{equation}
using a technique from Ref. \cite{aps}.  Here $T_{\ell \lambda}$
denotes any of the $E_\ell$ and $M_\ell$ operators.
The partial-wave expansion of
the current operator contains the photon energy only in multiplicative
factors on each term of the expansion.  The integration over all
coordinates except $x$ can therefore be calculated once for
each partial wave by the Monte Carlo code, and the result can then be
used to compute the full integral for as many energies as desired by
recomputing $\phi_{\alpha d}$ only, greatly reducing the amount of
computation.

\begin{figure}
\centerline{\epsfig{file=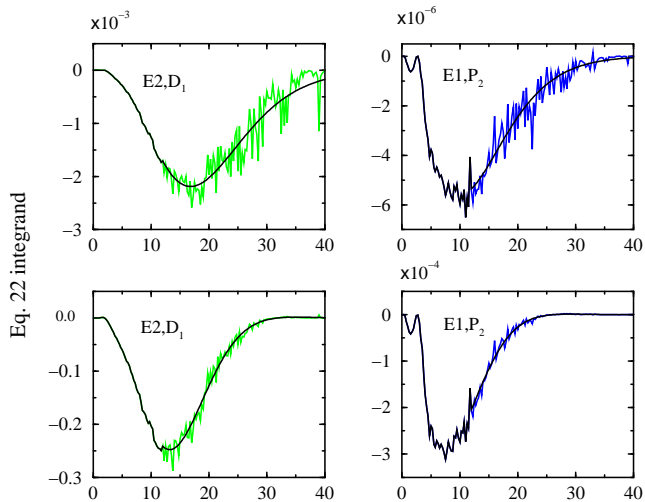,height=8.6cm,angle=270}}
\caption{Upper panel: The integrand of
Eq.(\protect\ref{eqn:integrand}) for $E2$ transitions from the $D_1$
initial state and $E1$ transitions from the $P_2$ initial state.  The
upper panels are for $E_{c.m.}=50$ keV, while the lower panels are for
$E_{c.m.}=1$ MeV.  ``Noisy'' curves are computed solely from the Monte
Carlo integration (and their scatter indicates sampling uncertainty),
while smooth curves show the expected long-range behavior of the
operators, with normalization fitted to the Monte Carlo results.}
\label{fig:integrand}
\end{figure}

The calculation sampled 2 000 000 Monte Carlo configurations, summing
over all 15 partitions of the six particles into clusters at each
step.  Typical results for the integrand in Eq.(\ref{eqn:integrand})
at low energy are shown for two transitions, at two different
energies, in Fig. \ref{fig:integrand}.  Calculations at energies above
about 1 MeV had integrands that were essentially zero beyond 30 fm
cluster separation, and therefore suffered little from limited
sampling.

We also performed a second set of calculations, this time computing
functions which had the expected asymptotic forms of the operator
densities as a function of $x$ at large cluster separations, and
normalizing them to the Monte-Carlo sampled operator densities between
12 and 22 fm.  These asymptotic forms, given by multiplying
Eq. (\ref{eqn:asymptotic}) by powers of $r_{\alpha d}$ appropriate for
each operator, were then used instead of the Monte Carlo operator
densities at cluster separations greater than 12 fm, in order to check
that our results were not affected by the effective cutoff in sampling
beyond 45 fm (due to very small $W({\bf R})$) or by poor sampling
elsewhere in the tails of the wave function.  For all of the dominant
reaction mechanisms, this gave essentially the same result as the
explicit calculation.  The smooth curves of Fig. \ref{fig:integrand}
were computed in this way.

\section{Results}
\label{sec:results}

\begin{figure}
\centerline{\epsfig{file=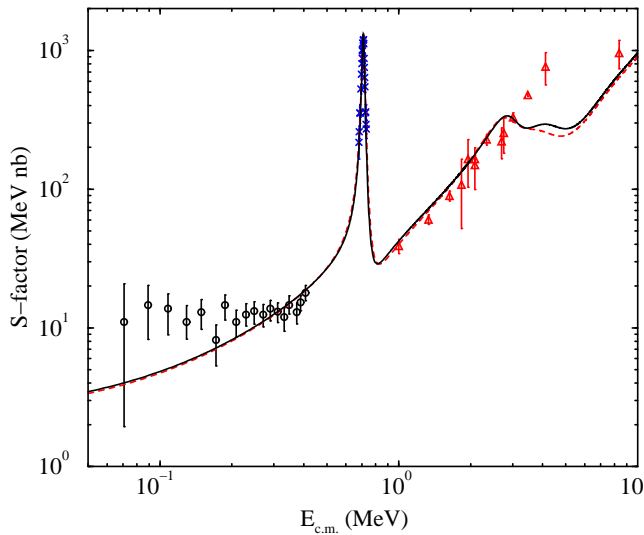,height=8.6cm,angle=270}}
\caption{The total $S$-factor calculated from the Langanke (solid
curve) and KP II potentials (dashed curve), with the data of
\protect\cite{rghr} ($\triangle$), \protect\cite{mohr} ($\times$), and
\protect\cite{kiener} ($\circ$).  The indirect data of
Ref. \protect\cite{kiener} were taken from their graphical
presentation in \protect\cite{nacre}.}
\label{fig:crsec}
\end{figure}

\begin{figure}
\centerline{\epsfig{file=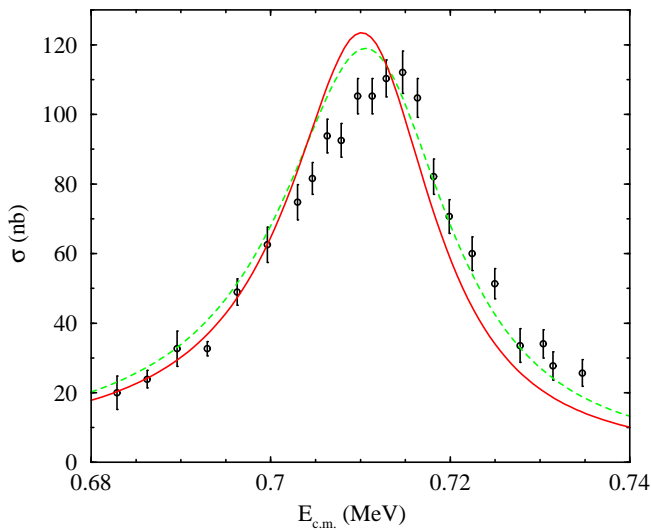,height=8.6cm,angle=270}}
\caption{The calculated total cross section at the 711 keV resonance,
compared with the data of Ref. \protect\cite{mohr}.  The solid curve
was computed from the Langanke potential, while the dashed curve was computed
from the KP II potential.}
\label{fig:peak}
\end{figure}

Our main result is the total capture cross section as a function of
energy, shown in comparison with laboratory data in
Fig. \ref{fig:crsec} as an astrophysical $S$-factor.  Note the good
agreement with the capture data below 3 MeV, reflecting a reasonable
value of the alpha-deuteron asymptotic normalization coefficient of
the $^6$Li ground state, $C_0$, and accurate reproduction of the
scattering phase shifts.  The disagreement with the data above 3 MeV
is a generic feature of direct-capture models, and is usually
attributed to neglected couplings to breakup channels, and to channels
with nonzero isospin in the scattering state (such as the $J^\pi,T=0^+,1$ 
state at 2.09 MeV).  As shown in Fig. \ref{fig:peak}, the
behavior at the $3^+$ resonance is nicely reproduced in location,
width, and amplitude, simply by constraining the cluster potentials in
the $D$ waves to reproduce the experimental phase shifts.  The failure
to match the energy dependence of the cross sections inferred below
500 keV from Coulomb-breakup experiments is shared by all other direct
capture calculations, and could reflect either something that is not
understood about the experiments, or something that is missing from
the models.  Our result with more detailed wave functions and
operators than previously considered may be taken as a further
indication that previous models were correct in this region.

\begin{figure}
\centerline{\epsfig{file=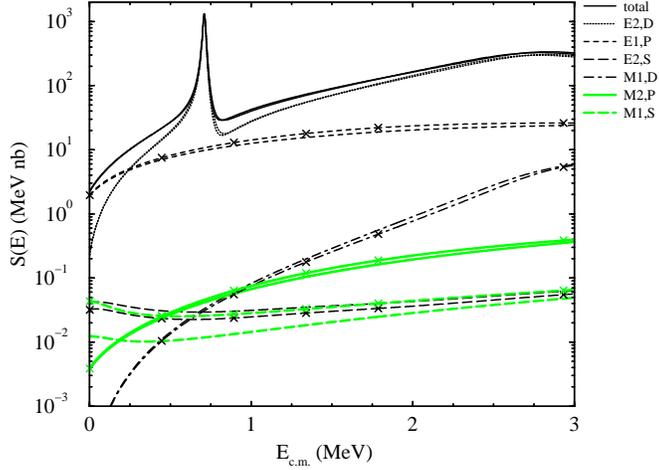,height=8.6cm
,angle=270}}
\caption{Contributions of the more important reaction multipolarities
and incoming partial waves to the $S$ factor.  Curves marked with
the symbols $\times$ were computed from the Langanke effective
potential; plain curves were computed from the KP II potential.  Curve
labels refer to reaction multipolarity and orbital angular momentum of
the initial state.}
\label{fig:breakdown}
\end{figure}

\begin{figure}
\centerline{\epsfig{file=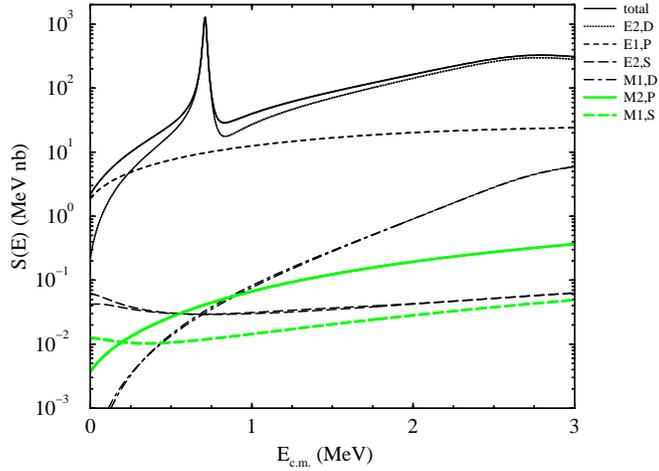,height=8.6cm,angle=270}}
\caption{Contributions of reaction mechanisms and incoming partial
waves to the $S$ factor, computed from the KP II potential.  Cross
sections were computed both using the plain Monte Carlo calculation,
and normalizing the expected asymptotic form to the Monte Carlo
operator densities; there are therefore two curves of each type,
although they are indistinguishable for the total and for some of the
mechanisms.  Curve labels are as in Fig.
\protect\ref{fig:breakdown}.}
\label{fig:breakdown-low}
\end{figure}

\begin{figure}
\centerline{\epsfig{file=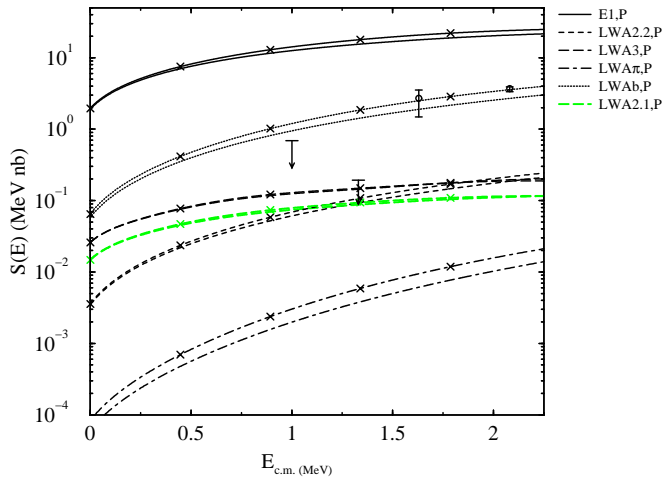,height=8.6cm,angle=270}}
\caption{The total $E1$ operator (solid curves) and the cross sections
resulting from each of the smaller terms of the operator acting
individually, shown as $S$ factors.  Plain curves were computed from
the KP II potential.  Curves with symbols $\times$ were computed from
the Langanke potential.  Labels for multipole and orbital angular
momentum are as in Fig. \protect\ref{fig:breakdown}.  Other labels are
as in Eqs. (\protect\ref{eq:lwcb}--\protect\ref{eq:lw33}), except that
LWA2.1 and LWA2.2 denote the first and second terms of
Eq. \protect\ref{eq:lw22}, respectively.  The data shown are the $E1$ cross
sections and upper limits from Ref.  \protect\cite{rghr}.}
\label{fig:e1}
\end{figure}

In Figs.~\ref{fig:breakdown} and \ref{fig:breakdown-low}, we show the
breakdown of the cross section into contributions from different
multipolarities and incoming partial waves, again as $S$ factors.
Fig.~\ref{fig:breakdown} shows results of two calculations, one from
the Langanke potential and one from the KP II potential.  The results
are dominated by $E2$ transitions arising from $D$-wave scattering and
by $E1$ transitions arising from $P$-wave scattering.  Results from
the two sets of scattering states are almost indistinguishable for
these transitions, especially at low energies.  In
Fig.~\ref{fig:breakdown-low}, the two sets of curves were both
computed from the KP II potential.  One set of curves was computed
directly from the Monte Carlo calculation, while the other was
computed from the expected asymptotic behavior of the integrand of
Eq.~\ref{eqn:integrand} beyond 12 fm, as described above.
Fig.~\ref{fig:e1} shows a further breakdown of the $E1$ operator into
its various terms, along with the measured $E1$ contributions
\cite{rghr}.  It is seen that:

1.) The $E1$ cross section is larger than the data of Robertson {\it
et al.} \cite{rghr} by a factor of about 7 at 2 MeV.  Most other
models have also arrived at $E1$ cross sections that are too large
relative to these data.  A new aspect of our calculation is that the
largest contribution is calculated from the nucleon-nucleon potentials
as a relativistic center-of-mass correction.  This corresponds closely
to inserting measured cluster masses by hand in simpler models, as
indicated by the fact that virtually identical (within 50\%) results
for the LWAc term are obtained from our six-body model and from a
simple capture model using the same cluster potential, but with
laboratory alpha and deuteron masses introduced.  In
Fig. \ref{fig:e1}, the top curve is the total $E1$ $S$ factor (arising
mainly from the center-of-energy correction), and the other curves
indicate the $S$ factors that would result from various small terms in
the E1 operator individually.  (These differ from their actual
contributions to the $S$ factor, since these contributions add
coherently in the matrix element.)  The center-of-energy and LWAb
terms are obviously much more important than the other terms in our
calculation, and the smaller terms do not affect the total
significantly.

We note also that Robertson {\it et al.} \cite{rghr} report the
opposite sign from what is expected for the $E1$ transition in their
experiment.  None of the effects in our calculation can produce an
effect of the correct magnitude with the observed ``wrong'' sign, or
reproduce the sharp drop in $E1$ amplitude from 1.63 to 1.33 MeV seen
in the data.

2.)  There is no significant $M1$ transition originating from the
$S$-wave scattering state at lowest order in LWA, and we estimate the
next-to-lowest term to be much smaller than what we obtain for the
lowest-order term.  The $M1$ strength is expected to be generally very
small, since it is zero by orthogonality arguments in a two-body
model.  However, at very low energy, any $S$-wave contribution present
may be expected to dominate the total cross section, and many-body
wave functions can (in principle) provide small pieces of the
final-state wave function which can be reached in such a transition.
Such an effect is present in neutron captures on the deuteron and on
$^3$He.  The presence of a nonzero $M1$ contribution in our
calculation raises two questions: to what extent is it a product of
our choice of wave functions, and how will the presence of meson
exchange currents (MEC), which is a factor-of-ten correction for the
$M1$ cross section in $^3{\rm He}(n,\gamma)^4{\rm He}$, affect our
result?  The answer to the first question lies in the discussion of
orthogonalization at the end of Sec. \ref{sec:wave functions} and in
Fig.  \ref{fig:breakdown}; orthogonal ground and scattering states
produce much smaller $M1$ contributions than non-orthogonalized
states, and the remaining $M1$ contribution is very sensitive to the
choice of phenomenological scattering potentials, suggesting a
near-cancellation that ought to be exact.  The second question is less
important than it may appear at first glance.  The large relative MEC
contributions to neutron captures on light nuclei typically arise from
the {\it isovector} part of the $M1$ operator.  The $\alpha d$ capture
process is isoscalar; the isoscalar MEC operator is both significantly
smaller than the corresponding isovector operator, and quantitatively
less certain in magnitude.  We have therefore neglected MEC
contributions to the M1 operator.  Only electric multipoles contribute
to Coulomb breakup, so the discrepancy between our results and the
indirect data (circles in Fig. \ref{fig:crsec}) is not the result of
omitting MEC.

To compare our results with those obtained in a simpler model, we have
also computed $\alpha d$ capture by treating the two clusters as point
particles, and the ground state as an $S$-wave bound state of these
two clusters.  We used the modified Langanke potential, described
above, to compute both states.  With this potential, the \li6 ground
state energy is 0.07 MeV too high, and this affects the energy
dependence of the cross section.  The asymptotic normalization of the
wave function, which sets the scale for the nonresonant cross section,
depends in any model on the detailed short-range behavior of the
ground state, and is larger for this model than for our six-body \li6
ground state.  While the 711 keV resonance is equally well-reproduced
in strength and width by both two-body and six-body models, the total
cross sections away from resonance are as much as 16\% smaller in the
six-body model.  The simple model also yields an $E1$ strength which
is greater by about 30\% at 3 MeV.  The source of the greater $E1$
strength is an interplay between the differing asymptotic
normalizations, the treatment of the center-of-energy correction
(treated in the simple model by using laboratory values of alpha and
deuteron masses), and differences in the sizes of LWA2 and LWAb terms.

\begin{figure}
\centerline{\epsfig{file=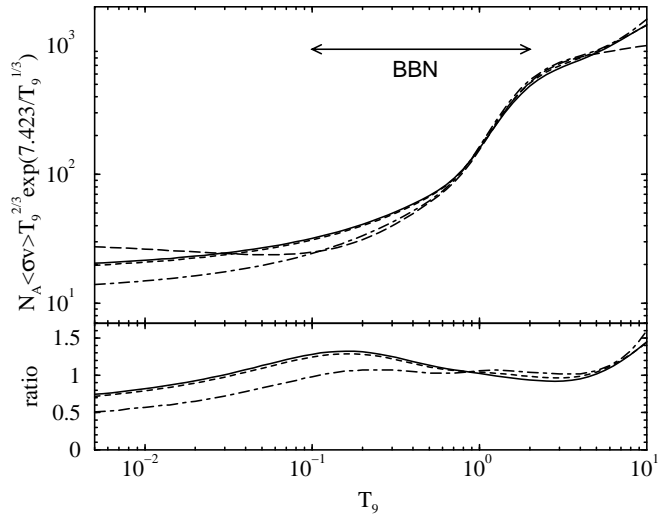,height=8.6cm,angle=270}}
\caption{Thermally-averaged reaction rates, for the Langanke (solid)
and KP II (dashed) potentials.  The upper panel shows rates scaled to
the temperature dependence expected for a constant $S$ factor (to make
small differences visible), while the lower panel shows ratios to the
rate of Robertson {\it et al.}  \protect\cite{rghr}.  Shown for
comparison are the Robertson {\it et al.} rate (long-dashed), and the
rate presently recommended in the NACRE compilation of reaction rates
(dash-dotted) \protect\cite{nacre}.  The line at the top indicates the
approximate temperature range for big-bang production of \li6.}
\label{fig:rate}
\end{figure}

We have also calculated thermally-averaged reaction rates suitable for
use in astrophysical calculations.  We integrate from 0.1 keV to 10
MeV at temperatures from $10^7$ K to $3\times 10^{10}$ K to produce
the quantity $N_A\langle\sigma v\rangle$ customarily used in reaction
network calculations.  Our result, shown in Fig. \ref{fig:rate}, is as
much as 40\% larger than previous estimates over a wide range of
temperatures.  The sources of these differences are not clear.  They
probably arise from several specific details of the capture models
used, which have been normalized to the data using assumptions about
the relative roles of the $E1$ and $E2$ components to produce the
rates recommended in compilations.  

\section{Discussion}
\label{sec:end}

The total cross sections we have computed compare favorably with the
data in the region from the $3^+$ resonance to 3 MeV.  This reflects
mainly a combination of reasonable fits of the scattering states to
experimental phase shifts, and reasonable reproduction of the
asymptotic behavior of the $^6$Li ground state in the $S$-wave $\alpha
d$ channel.  Our calculation of the $E1$ cross section from the
relativistic center-of-energy correction, and examination of several
smaller corrections, is new; however, like all other models not
actually scaled to the laboratory data, it is higher than the data.
In our calculation, the $E1$ transition contributes just under 70\% of
the total cross section at 50 keV.

A very large cross section for alpha-deuteron capture would be of some
interest for cosmology, since it would allow significant production of
\li6 in the big bang.  However, we find that big-bang nucleosynthesis
calculations using our new cross sections can produce \li6 at a
maximum level of 0.000 60 relative to $^7$Li, or \li6/H=$5\times
10^{-14}$.  The maximum occurs at a baryon/photon ratio of $2\times
10^{-10}$, at the low end of the range of possible values suggested by
other light-element abundances.  At the higher baryon/photon ratio of
$5\times 10^{-10}$ suggested by recent extragalactic deuterium
abundance measurements, there is a factor of three less \li6, but
because of the increasing $^7$Li production, \li6/$^7$Li = 0.000 07.
Since the abundance ratio found in low-metallicity halo stars so far
is in the vicinity of \li6/$^7$Li = 0.05, it seems very unlikely that
standard big-bang nucleosynthesis could account for observed
abundances.  It should be kept in mind that non-standard models not
relying solely on the alpha-deuteron capture mechanism could, in
principle, produce interesting amounts of \li6 \cite{jedamzik}.

In the future we expect to make several improvements on the present work.
While the VMC wave functions for the s-shell nuclei give energies within 2\% 
of the experimental value, the trial functions for p-shell nuclei are not 
as good.  It should be worthwhile to try the more accurate GFMC ground states 
for the p-shell nuclei and calculate mixed estimates of the transition 
operators~\cite{PPCPW97}.  Eventually one would also like to construct
the scattering states with the GFMC method, but this will be
considerably more difficult.
Further, new three-body potentials that give a better fit to the energies of
the light p-shell nuclei are currently under development~\cite{PPRWC00}
and should improve both the VMC and GFMC results.
An important aspect of these studies will continue to be the construction of
ground states that have the proper asymptotic cluster behavior.  This must
be imposed in the variational trial function, because the GFMC algorithm
is not sensitive to the tail of the wave function and will not be able to
correct for any failures.

The present calculation should serve as a useful starting point for studies
of several other reactions of astrophysical interest involving p-shell nuclei, 
including $^4{\rm He}(t,\gamma)^7{\rm Li}$, 
$^4{\rm He}(^3{\rm He},\gamma)^7{\rm Be}$,
and $^7{\rm Be}(p,\gamma)^8{\rm B}$.  
Work on these reactions is now in progress.

\acknowledgments

The authors wish to thank V.R.\ Pandharipande and S.C.\ Pieper 
for many useful comments.
Computations were performed on the IBM SP of the Mathematics and Computer
Science Division, Argonne National Laboratory.
The work of KMN and RBW is supported by the U. S. Department of Energy, Nuclear
Physics Division, under contract No. W-31-109-ENG-38, and that of RS
by the U. S. Department of Energy under contract No. DE-AC05-84ER40150.

%\bibliographystyle{prsty} 
%\bibliography{li6}

\end{document}